\documentclass[aps,prl,draft,showpacs,twocolumn]{revtex4}

\newcommand{\ket}[1]{\left | #1 \right \rangle}

\begin{document}
\title{Comment on ``Geometric phase of entangled spin pairs in a magnetic field''}
\author{Yang Xiang}
\email{njuxy@sina.com}
\author{  Shi-Jie Xiong }

\affiliation{National Laboratory of Solid State Microstructures and
Department of Physics, \\Nanjing University, Nanjing 210093, China}
\date{\today}
\begin{abstract}
The degree of entanglement between two spins may change due to
interaction. About this we find that a wrong result in a recent work
by Ge and Wadati [Phys. Rev. A {\bf72}, 052101(2005)] which breach
the basic principle.
\end{abstract}

 \maketitle

Recently, Ge and Wadati \cite{ge} studied the geometric phase in the
case of two initially entangled interacting $\frac{1}{2}$-spins
under a time-independent magnetic field. The Hamiltonian operator of
the model is
\begin{equation}
H= \sum_{i=1,2}\omega_{i}S^{(i)}_{z}+\frac{8\lambda}{\hbar}
\sum_{q=x,y,z}a_{q}S^{(1)}_{q}S^{(2)}_{q}, \label{hamilton}
\end{equation}
where ${\bf S^{(1)}}=(S^{(1)}_{x},S^{(1)}_{y},S^{(1)}_{z})$ and
${\bf S^{(2)}}=(S^{(2)}_{x},S^{(2)}_{y},S^{(2)}_{z})$ are the spin
operators of the two particles, $\omega_{1}$ and $\omega_{2}$ are
the corresponding Larmor frequencies originating from a
time-independent magnetic field along the $z$ axis, $\lambda$ and
$(a_{x},a_{y},a_{z})$ are the strength and anisotropy of the
interaction. There is an interaction between the two spins, so the
degree of entanglement may change due to the interaction. But we
find that Ge and Wadati \cite{ge} have ignored this.

Firstly, let us consider a $\frac{1}{2}$-spin precessing in a
time-independent uniform magnetic field pointing in the $z$
direction. Assuming that the angle between the spin of the initial
state $\ket{{\bf n}(t=0)}$ and the $z$ axis is $\theta$, the state
at a later time $t$ is
\begin{eqnarray}
\ket{{\bf n}(t)}=e^{-i\varphi(t)/2}\cos\left(\frac{\theta}{2}\right)
\ket{+z}+e^{i\varphi(t)/2}\sin\left(\frac{\theta}{2}\right)\ket{-z}.
\label{one state}
\end{eqnarray}
Here $\varphi(t)=\varphi(0)+\omega t$ with $\varphi(0)$ being the
initial azimuth angle with the $x$ axis and $\omega$ is the Larmor
frequency. At the same time, any pure state of two
$\frac{1}{2}$-spins can be decomposed as
\begin{eqnarray}
\ket{\Psi}&=&e^{-i\beta/2}\cos(\frac{\alpha}{2})\ket{{\bf
n}}_{1}\ket{{\bf
m}}_{2}\nonumber\\
&~~~~&+e^{i\beta/2}\sin(\frac{\alpha}{2})\ket{{\bf -n}}_{1}\ket{{\bf
-m}}_{2} \label{initial state}
\end{eqnarray}
according to the Schmidt theorem \cite{schmidt}, where ${\bf n}$ and
${\bf m}$ are two points on the Poincar\'{e} sphere, and the
subscripts denote spins $1$ and $2$ respectively. The ``angle''
$\alpha$ in Eq. (\ref{initial state}) determines the degree of
entanglement in the state. For a bipartite pure state, the measure
of entanglement is the von Neumann entropy of the reduced state in
either of the two parties \cite{bennett}. For the state $\ket{\Psi}$
in Eq. (\ref{initial state}) we can calculate its entanglement
$E(\ket{\Psi})$ as
\begin{eqnarray}
E(\ket{\Psi})&=&-\cos^{2}(\frac{\alpha}{2})\log_{2}{
\cos^{2}(\frac{\alpha}{2})}\nonumber\\
&~~~~&-\sin^{2}(\frac{\alpha}{2})\log_{2}{\sin^{2}(\frac{\alpha}{2})}.
\end{eqnarray}

In \cite{ge}, Ge and Wadati assumed that the initial state of two
spin-$\frac{1}{2}$ particles is
\begin{eqnarray}
\ket{\Psi(0)}&=&e^{-i\beta/2}\cos(\frac{\alpha}{2})\ket{{\bf
n(0)}}_{1}\ket{{\bf
m(0)}}_{2}\nonumber\\
&~~~~&+e^{i\beta/2}\sin(\frac{\alpha}{2})\ket{{\bf
-n(0)}}_{1}\ket{{\bf -m(0)}}_{2}, \label{initial state1}
\end{eqnarray}
where $\ket{{\bf n}(0)}$ and $\ket{{\bf m}(0)}$ are
\begin{eqnarray}
\ket{{\bf
n}(0)}=e^{-i\varphi_{1}/2}\cos\left(\frac{\theta_{1}}{2}\right)\ket{+z}
+e^{i\varphi_{1}/2}\sin\left(\frac{\theta_{1}}{2}\right)\ket{-z}\nonumber
\end{eqnarray}
and
\begin{eqnarray}
\ket{{\bf
m}(0)}=e^{-i\varphi_{2}/2}\cos\left(\frac{\theta_{2}}{2}\right)\ket{+z}
+e^{i\varphi_{2}/2}\sin\left(\frac{\theta_{2}}{2}\right)\ket{-z}\nonumber
\end{eqnarray}
(see Eq.(12) and Eq.(13) in \cite{ge}), then they argued that by
applying $U(t)=e^{-i H t}$ on $\ket{\Psi(0)}$ they can obtain
\begin{eqnarray}
\ket{\Psi(t)}&=&e^{-i\beta/2}\cos(\frac{\alpha}{2})\ket{{\bf
n(t)}}_{1}\ket{{\bf
m(t)}}_{2}\nonumber\\
&~~~~&+e^{i\beta/2}\sin(\frac{\alpha}{2})\ket{{\bf
-n(t)}}_{1}\ket{{\bf -m(t)}}_{2} \label{final state}
\end{eqnarray}
(see Eq. (14) in \cite{ge}), where $\ket{{\bf n}(t)}$ and $\ket{{\bf
m}(t)}$ are given by Eq. (\ref{one state}), and $H$ is expressed in
Eq. (\ref{hamilton}). It is noted that $\ket{\pm {\bf n}(t)}$ are
always orthogonal with each other. As the ``angle'' $\alpha$ in the
Schimdt decomposition determines the degree of entanglement in the
state, from the comparison of Eq. (\ref{initial state1}) with Eq.
(\ref{final state}), one can easily find that
\begin{eqnarray}
E(\ket{\Psi(0)})=E(\ket{\Psi(t)}),
\end{eqnarray}
i.e., the entanglement of $\ket{\Psi(0)}$ equals the entanglement of
$\ket{\Psi(t)}$. However, this is not true when there exists an
interaction between the two spins.

In fact, Sj\"{o}qvist \cite{sjoqvist} has studied the geometric
phase of a closed quantal system consisting of two
spin-$\frac{1}{2}$ particles with a spin-spin interaction. The
Hamiltonian operator is
\begin{eqnarray}
H=\frac{2\lambda}{\hbar}{\bf S}^{(1)}\cdot{\bf S}^{(2)},
\label{hamilton2}
\end{eqnarray}
similar to the interaction term in Eq. (\ref{hamilton}). In this
work Sj\"{o}qvist pointed out that the degree of entanglement may
change due to the interaction between the two spins, a product state
may evolve into an entangled state and vice versa, and it is
impossible to make one of the subsystems evolve but to keep the
other unchanged. A simple example has been proposed to illustrate
this in \cite{sjoqvist}. For two spin-$\frac{1}{2}$ particles under
Hamiltonian of Eq. (\ref{hamilton2}), if the initial state is
\begin{eqnarray}
\ket{\Psi(0)}=\cos\left( \frac{a}{2}\right)\ket{1,0}+\sin\left(
\frac{a}{2} \right)\ket{0,0}, \label{1}
\end{eqnarray}
where
$\ket{1,0}=\frac{1}{\sqrt{2}}(\ket{+z}_{1}\ket{-z}_{2}+\ket{-z}_{1}\ket{+z}_{2})$
and
$\ket{0,0}=\frac{1}{\sqrt{2}}(\ket{+z}_{1}\ket{-z}_{2}-\ket{-z}_{1}\ket{+z}_{2})$,
with subscripts denoting spin $1$ and $2$, we can express
$\ket{\Psi(0)}$ in terms of the Schmidt decomposition
\begin{eqnarray}
\ket{\Psi(0)}=\cos\left( \frac{\alpha}{2} \right)
\ket{+z}_{1}\ket{-z}_{2}+\sin \left( \frac{\alpha}{2} \right)
\ket{-z}_{1}\ket{+z}_{2}, \label{2}
\end{eqnarray}
where $\alpha$ satisfies
$\tan(\frac{\alpha}{2})=(\cos(\frac{a}{2})-\sin(\frac{a}{2}))
/(\cos(\frac{a}{2})+\sin(\frac{a}{2}))$. This means that if the
initial state is given, $\alpha$ is also given as a parameter. The
value of $\alpha$, however, may evolve in time under the
interaction. By applying $U(t)=e^{-i H t}$ on $\ket{\Psi(0)}$, we
can obtain $\ket{\Psi(t)}$ as
\begin{eqnarray}
\ket{\Psi(t)}&=&e^{-i \lambda t/2}\cos\frac{a}{2}\ket{1,0}+e^{i 3
\lambda t/2}\sin\frac{a}{2}\ket{0,0}\nonumber\\
&=&\frac{e^{i \lambda t/2}}{\sqrt{2}}[\left(e^{-i \lambda
t}\cos\frac{a}{2}+e^{i \lambda
t}\sin\frac{a}{2}\right)\ket{+z}_{1}\ket{-z}_{2}\nonumber\\
&~~~~&+\left(e^{-i \lambda t}\cos\frac{a}{2}-e^{i \lambda
t}\sin\frac{a}{2}\right)\ket{-z}_{1}\ket{+z}_{2}].
\end{eqnarray}
Ignoring an overall phase we have
\begin{eqnarray}
\ket{\Psi(t)}&=&e^{-i
\beta(t)/2}\cos\frac{\alpha(t)}{2}\ket{+z}_{1}\ket{-z}_{2}\nonumber\\
&~~~~&+e^{i
\beta(t)/2}\sin\frac{\alpha(t)}{2}\ket{-z}_{1}\ket{+z}_{2},
\end{eqnarray}
where $\cos\alpha(t)=\sin a\cos2\lambda t$ and $\tan\beta(t)=-\tan
a\sin 2 \lambda t$. The ``angle'' $\alpha(t)$ is changing with time.
This means that the degree of entanglement is also changing with
time.

To summarize, we point out that there is a mistake in \cite{ge}
which breaks a basic principle, i.e., the degree of entanglement may
change due to the interaction between two spins. We illustrate this
by using a simple example.





%
%
%
%
%
%

\vskip 0.5 cm

{\it Acknowledgments} This work was supported by the State Key
Programs for Basic Research of China (2005CB623605 and
2006CB921803), and by National Foundation of Natural Science in
China Grant Nos. 10474033 and 60676056.


%

\end{document}